\documentclass[sigconf]{acmart}
\usepackage[export]{adjustbox}
\usepackage{amsmath}
\usepackage{amsfonts}
\usepackage{amsthm}
\usepackage{algpseudocode}
\usepackage{algorithm}
\usepackage{rotating}
\usepackage{booktabs}
\usepackage{multirow}
\usepackage{graphicx}
\usepackage{xspace}
\usepackage{enumitem}
\usepackage{subfigure}

\newcommand{\eg}{\textit{e.g.}}
\newcommand{\ie}{\textit{i.e.}}
\newcommand{\etc}{\textit{etc}}
\newcommand{\wrt}{\textit{w.r.t.}}

\newcommand{\nosection}[1]{\vspace{2pt}\noindent\textbf{#1}}

\AtBeginDocument{%
  \providecommand\BibTeX{{%
    \normalfont B\kern-0.5em{\scshape i\kern-0.25em b}\kern-0.8em\TeX}}}

\setcopyright{acmlicensed}
\copyrightyear{2018}
\acmYear{2018}
\acmDOI{XXXXXXX.XXXXXXX}

\acmConference[Conference acronym 'XX]{Make sure to enter the correct
  conference title from your rights confirmation emai}{June 03--05,
  2018}{Woodstock, NY}
\acmISBN{978-1-4503-XXXX-X/18/06}

\begin{document}


% Reranking Search Engine System with Large Language Model on Short Video Platform
\title{LLM4PR: Improving Post-Ranking in Search Engine with Large Language Models}

\author{Yang Yan, Yihao Wang, Chi Zhang, Wenyuan Hou, Kang Pan, Xingkai Ren, Zelun Wu, Zhixin Zhai, Enyun Yu, Wenwu Ou and Yang Song}
% \author{Zhang Chi}
\affiliation{%
  \institution{Kuaishou Technology}
  % \streetaddress{P.O. Box 1212}
  \city{Beijing}
  \country{China}
}

% \affiliation{%

% \affiliation{%

% \affiliation{%

% \affiliation{%

% \affiliation{%

% \affiliation{%

% \affiliation{%

\renewcommand{\shortauthors}{Yan and Zhang, et al.}

\begin{abstract}

% Alongside the rapid development of Large Language Models (LLMs),  there has been a notable increase in efforts to integrate LLM techniques into search engines (SE). However, research dedicated to enhancing the re-ranking stage in SE through LLMs is still largely unexplored. In this study, we introduce a novel paradigm that harnesses the capabilities of Large Language Models for the Re-ranking Stage in search engines, which we refer to as LLM4PR. As far as we know, the proposed LLM4PR is the LLM-based framework studying the re-ranking problem in search engine. Particularly, the LLM4PR framework first suggests a Query-Instructed Adapter (QIA) module to align heterogeneous features including the semantic information and user/item features. The LLM4PR further introduces a  learning to re-rank step, where two task templates are suggested and the model is fune-tuned to adapt the re-ranking task. Both offline and online experiment studies demonstrate that the proposed framework leads to significant improvements and exhibits state-of-the-art performance  compared with other alternative.
% ======================================================
Alongside the rapid development of Large Language Models (LLMs),  there has been a notable increase in efforts to integrate LLM techniques in information retrieval (IR) and search engines (SE).  Recently, an additional ``post-ranking'' stage  is suggested in SE to enhance user satisfaction in practical applications. Nevertheless, %despite the critical role as the final stage in SE that generates the ultimate outcomes, 
research dedicated to enhancing the post-ranking stage through LLMs remains largely unexplored. 
In this study, we introduce a novel paradigm named Large Language Models for Post-Ranking in search engine (\textbf{LLM4PR}), which leverages the capabilities of LLMs to accomplish the post-ranking task in SE.
% As far as we know, the proposed LLM4PR is the first LLM-based framework focused on studying the re-ranking problem in search scenarios. 删掉
Concretely, a Query-Instructed Adapter (QIA) module is designed to derive the user/item representation vectors by incorporating their heterogeneous features.
A feature adaptation step is further introduced to align the semantics of user/item representations with the LLM.
% and a feature adaptation step to seamlessly incorporate  diverse user/item features into the LLM input and align the semantics of feature representations with the LLM. % 两句分开写
Finally, the LLM4PR integrates a learning to post-rank step, leveraging both a main task and an auxiliary task to fine-tune the model to adapt the post-ranking task. 
Experiment studies demonstrate that the proposed framework leads to significant improvements and exhibits state-of-the-art performance compared with other alternatives.
% ranking order 都替换掉，reranking order, final results.
% 图里的模板符号换一下

\iffalse
\textcolor{blue}{1. title: improving search post-ranking with LLM 2.no need to claim as far as we know. 3. add the definition of re-ranking before however 4. do not integrate QIA and FA in one sentence 5. figure 1: do not say final ranking order 6. figure 4, need to claim it is template 7.significant test. 8. stress listwise in paragraph 2}
\fi

%In addition, current LLMs for RS/SE neglects the mutual information between items. 

\end{abstract}





\maketitle

\section{Introduction}

\begin{figure}[ht]
    \centering
    \includegraphics[width=0.95\columnwidth]{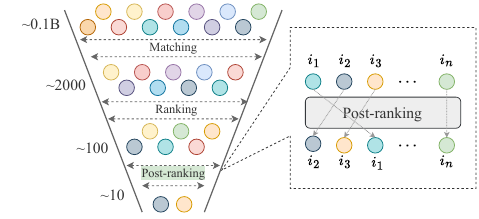}
    \vspace{-10pt}
    \caption{Illustration of the search process, encompassing  matching, ranking and post-ranking stages.}
    \label{fig:postranking_paradigms}
    % \vspace{-15pt}
\end{figure}

Conventional search engines (SE) and information retrieval (IR) systems are composed of multiple stages including matching and ranking, in which the relevant items are retrieved in matching stages and sorted by relevance in ranking stage. However, in practical applications, the relevance score of a single item is not the sole measure for practical search engines. On e-commerce platforms, the search engine is required not only to provide relevant results but also to maximize users' purchase rate by presenting the product list. Meanwhile, in the short video search scenario, the search platform also attempts to maximize users' play time during the entire search session. To this end, a post-ranking stage is suggested in practical applications. As depicted in Figure \ref{fig:postranking_paradigms}, the post-ranking stage severs as the final stage in search engines. This stage considers multiple attributes (item relevance score, user click/purchase rate, \etc.) as well as mutual influences between items, and deliver the final list to optimally enhances user experience in the search session. %through a listwise approach,  which takes into account multiple item attributes (relevance score, click/purchase rate, \etc.) as well as mutual influences between items. 

In IR and SE, the classic Learning-To-Rank (LTR) method formulates a composite objective encompassing item relevance and click/purchase to model the post-ranking stage. Recently, numerous additional efforts have been dedicated to studying this search topic\cite{liu2022ijcai, ai2019ictir,Zhuang2018midnn,Gong2020edgerec,Abdool2020airbnb, Pang2020setrank, Huzhang2023EGrank}. Concretely, SetRank \cite{Pang2020setrank} proposes the self-attention method to model mutual influences between items, and  DLCM \cite{ai2018sigir} suggests a RNN-based approach. PRM \cite{pei2019prm}, PRS \cite{feng2021prs} and PIER \cite{shi2023pier} suggest permutation-based two-stage processes for list generation and evaluation. Another pioneering stream of studies on post-ranking stages focus on the generative method. Seq2slate\cite{Bello2019seq} and GRN \cite{feng2021grn}  introduce the generative method in post-ranking stage in recommendation scenario\footnote{In recommender system, ``re-ranking'' refers to ``post-ranking'' stage. While in information retrieval, "re-ranking" denotes the "ranking". In case of confusion, we call the last stage in search engine as ``post-ranking'' in this paper. }. Regarding the increasing popularity of generative methods, especially its significant success in Large Language Models (LLMs), the generative and LLM-based approaches emerge as a promising avenue for optimizing post-ranking stage in search engine.

The advances in Large Language Models (LLMs) have achieved remarkable success in Natural Language Processing (NLP) and IR tasks \cite{Vaswani2017trm, Devlin2019bert, Touvron2023llama, Brown2020gpt3, openai2023gpt4,touvron2023llama2}, and it has been prompting increasing integration between LLM and practical applications such as search engine (SE). Currently, most existing LLM methods for SE primarily focus on augmenting document retrieval \cite{Bonifacio2022inpars, Jeronymo2023inParsv2, Dai2023Promptagator, ma2023finetunellama, wang2024improvingtextllm}, document ranking/re-ranking \cite{qin2023llmpairranker, sun2023listwisereranker} and modeling relevance between queries and items  \cite{zhuang2023potintwisereranker, qin2023llmpairranker, Sachan2022pointwisereranker,cho2023pointwisereranker}. 
However, despite significant efforts being directed towards matching (\ie, item retrieval) and ranking within SE,  the post-ranking stage has been overlooked. Consequently, the explorations of  LLMs for the post-ranking stage in SE remain scarce. To implement LLMs for the post-ranking process in search engine, several challenges have yet to be overcome.

\textbf{Issue 1: Heterogeneous Features Input.} The input of post-ranking model commonly encompasses heterogeneous features including item descriptions, category ID features, item-level statistical features, and the output (\ie, numerical embedding) from the up-stream (\ie, ranking) stage. However, current LLMs are specifically designed to process semantic textual input only. Directly inputting these feature embeddings into the LLM could potentially lead to confusion, as these embeddings lack semantic context. 
%has the potential to cause confusion, since these embeddings lack semantic information. 
Therefore, developing a mechanism to seamlessly incorporate these diverse features into the LLM input and align the feature representation with the LLM remains a significant challenge. 

\textbf{Issue 2: Task Specification for Post-Ranking.} Existing LLMs are designed for general purposes (\ie conversation, question and answering), exhibiting limited capacities for the post-ranking task in practical applications. Thus, it is essential to explore appropriate adjustments to enhance the LLM's ability for optimizing the post-ranking task.

To address these issues, we propose an \underline{L}LM-based framework for  \underline{P}ost-\underline{R}anking in search engine, named \textbf{LLM4PR}, comprising a Query-Instructed Adapter (QIA) component and a backbone LLM. 
% In this manuscript, we propose an LLM for Search Re-ranking framework (LLM4PR), which consists of the Query-Instructed Adapter (QIA) component and LLM. 
To handle the challenge of heterogeneous features (\textbf{Issue 1}), we propose a feature adaptation step to integrate diverse features. The QIAs take user/item side features as input, and combine various features in a query-instructed manner. Specifically, within the QIAs, the search query assigns distinct weights to diverse feature domains via the attention mechanism and derive the representation vectors for the user/item. Moreover, a template-based generation task is designed to achieve the semantic alignment between the user/item representations and the LLM. Particularly, we freeze the LLM and train the QIAs to generate semantic user/item representations that can be decoded by the LLM.
% \ie, to generate an appropriate description for the input user/item representation vector. 
To tackle the post-ranking task challenge (\textbf{Issue 2}), we design a main task and an auxiliary task to fine-tune the \textbf{LLM4PR} to learn to post-rank, named learning to post-rank step. The main task involves taking candidate item list as input and instructing \textbf{LLM4PR} to predict the target post-ranking order in a generative manner. To produce satisfying post-ranking lists, the model is required to assess the quality of various candidate item lists.
% The requisite for generating an excellent ranking order is the ability to judge the quality of different ranking orders. 
To this end, we propose an auxiliary task that compares a pair of candidate lists and predicts the superior one. The auxiliary task serves as a supplement to the main task, supporting the model in developing an insight for judging the quality of candidate lists. In the training stage, both the main task and the auxiliary task are trained simultaneously. During the inference stage, we utilize the main task to generate the final post-ranking results.

Overall, the contributions of this paper can be summarized as follows:
%\vspace{-0.45cm}
\begin{itemize}[leftmargin=*]
% \begin{itemize}
    \item Firstly, to the best of our knowledge, our \textbf{LLM4PR} is the first LLM-based framework that optimizing the post-ranking stage for search engines. This framework leverages the robust capabilities of LLM and generates the final results straightforwardly.

    \item Secondly, the proposed \textbf{LLM4PR} suggests a novel QIA component and feature adaptation step to align heterogeneous features. Additionally, it introduces an efficient template-based learning to post-rank step for model tuning.

    % \item Secondly, the proposed LLM4PR suggests a novel QIA component to align heterogeneous features, and introduces an efficient task-template based learning to re-rank step for model tuning.
 
    \item Last but not least, the effectiveness of the proposed \textbf{LLM4PR} is fully validated through comprehensive experiment studies.
    % Particularly, going beyond offline experiments in typical LLM studies for search engines, our LLM4PR has been deployed on Kuaishou short video platform and demonstrated substantial online performance improvements.

\end{itemize}

% The merits of our proposed xxx framework are following:

% adapter, query 和item交互
% aligh with ranking
%

\section{Related Works}
%since the introduction of the Transformer architecture \cite{Vaswani2017trm}. Recently, %benefit from the increasingly large corpus and strong hardware training power,

\nosection{Post-Raking in Search Engine.} 
In the context of the post-ranking stage, where not only document relevance but also user satisfaction are taken into consideration in practical applications,  \cite{liu2022ijcai} categorizes the methods from the perspectives of modeling objectives and learning signals. Early methods \cite{ai2018sigir, pei2019prm, Pang2020setrank, Zhuang2018midnn, shi2023pier, feng2021prs} consider the single objective and learn from supervised signals. \cite{Abdool2020airbnb} proposes location diversity ranker loss to optimize the diversity issue in Airbnb Search. Among generative post-ranking models, ListCVAE \cite{Jiang2019rerankcvae} selects the Conditional Variational AutoEncoder (CVAE) to generate the slate straightforwardly, while Seq2slate \cite{Bello2019seq} and GRN \cite{feng2021grn}  predict the next item sequentially via pointer network and policy gradient, respectively. \cite{Huzhang2023EGrank} further introduces a discriminator to ensure the generated results are close to real lists. 

\nosection{{LLM for Information Retrieval. }}
In recent years, research on Large Language Models (LLMs) has gained considerable attention. Various pre-trained models (BERT\cite{Devlin2019bert}, T5\cite{Vaswani2017trm, Raffel2020t5}, BART\cite{Lewis2020Bart}) and LLMs (LLaMA series \cite{Touvron2023llama,touvron2023llama2}, GPT series \cite{Brown2020gpt3, openai2023gpt4}, Baichuan \cite{yang2023baichuan}, GLM \cite{du2022glm}) are proposed, and \cite{zhao2023llmsurvey} provides a comprehensive survey reviewing the development of LLMs. 
In light of the remarkable successes achieved by LLMs for natural language problems, considerable efforts have been dedicated to integrating LLMs with information retrieval. \cite{zhu2023llmLR} offers an insightful overview of potential applications of LLMs in information retrieval, including query rewriter, retriever, text ranking and so on. The text ranking capabilities of LLMs have been comprehensively explored through various methodologies, including the pointwise approach \cite{zhuang2023potintwisereranker, Sachan2022pointwisereranker,cho2023pointwisereranker}, the pairwise approach \cite{qin2023llmpairranker}, and the listwise approach \cite{sun2023listwisereranker, ma2023listwiserereranker}. \cite{ma2023finetunellama} proposes a LLaMA-based retriever and ranker simultaneously. \cite{sun2023listwisereranker} investigates ChatGPT for passage ranking straightforwardly and suggests model distillation. Further works on LLMs for document ranking could be found in \cite{Wu22023llmrecsurvey,liao2023llara, Harte2023llmseq, li2023e4srec, qiu2023controlrec}. However, the aforementioned methods do not pay attention to the post-ranking stage, and LLM-based methods for post-ranking in search engine are urgently needed. 

%\cite{qin2023llmpairranker} presents a pairwise ranking prompt for document ranking. 

%In addition to document ranking,  the ranking ability of LLMs is also investigated in recommender systems. The related survey and works could be found in \cite{Wu22023llmrecsurvey,liao2023llara, Harte2023llmseq, li2023e4srec, qiu2023controlrec}.\wyh{However, the aforementioned methods do not  on optimizing post-ranking stage, they all neglect its importance in information retrieval and retrieval search engines.}

\begin{figure*}[ht]
    \centering
    \includegraphics[width=0.90\textwidth]{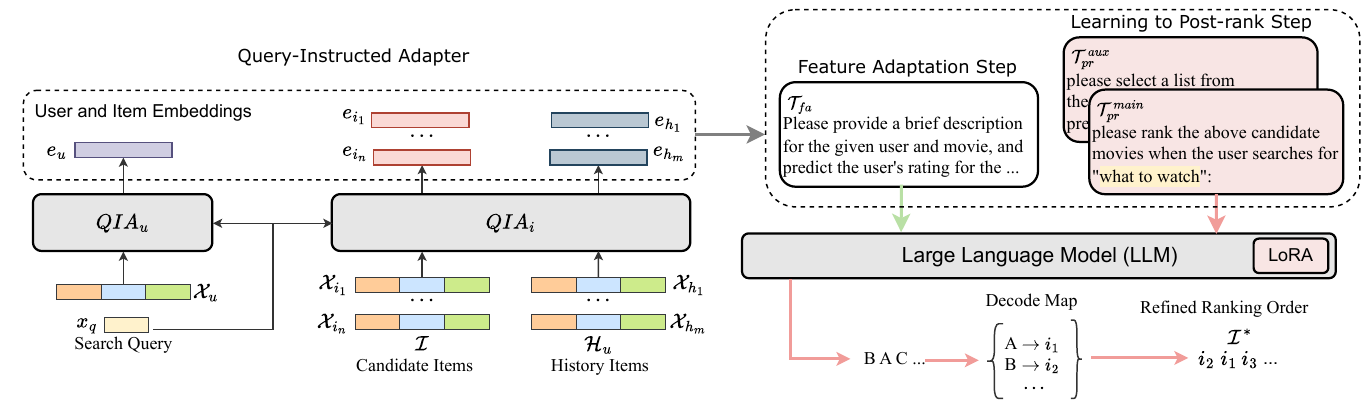}
    \vspace{-10pt}
    \caption{Overview of the proposed LLM4PR framework. } 
    \label{fig:model_structure}
    % \vspace{-10pt}
\end{figure*}

% The model takes query embedding $x_q$, user profile features $\mathcal{X}_u$, candidate item features $\mathcal{X}_i$ and user behavior items $\mathcal{X}_h$ information as input, and generates the final list $\mathcal{I}^{\ast}$ directly.

{\section{Preliminary}}
%Re-ranking is the last stage in search engine, which tailors the results list by considering mutual influences between items and determines the final results presented to the users. 
In search engines, post-ranking refines the results list by accounting for both item quality and the mutual influences between items, and determines the ultimate outcomes to maximize users experience. Particularly, given a request $\mathcal{R}$ and its output list $\mathcal{I}=\{i_1,i_2,\cdots,i_n\}$ from previous stage (\ie, ranking stage), the post-ranking algorithm aims to refine the order and produce a new ranking list $\mathcal{I^{\ast}}=\{i_1^{\ast},i_2^{\ast},\cdots,i_n^{\ast}\}$ that better matches the users preferences. Therefore, the post-ranking stage can be formulated as a function $f:\mathcal{R}\times\mathcal{I}\mapsto\mathcal{I^{\ast}}$, where the request $\mathcal{R}=\{q,\mathcal{X}_u,\mathcal{H}_u\}$ comprises the raw search query $q$, user profile features $\mathcal{X}_u$, and information about user history behavior items $\mathcal{H}_u$.

% \section{Reranking For Search Engine System }
\section{Methodology}
% The key idea of our proposed LLM4PR is constructing the final ordered list through a generative approach. In this section, firstly, we present the overview structure of the proposed LLM4PR and describe the generation procedure for re-ranking. Secondly, we elaborate on the integrated Query-Instructed Adapter (QIA), designed to generate appropriate user/item embeddings for LLM. We further design the  feature adaptation step and learning to re-rank step, which respectively bridge the gap in embedding distribution and endow LLM4PR with re-ranking capacity. Finally, we discuss the training and inference process.

% \subsection{Reranking as Generation}

\subsection{Overview of LLM4PR}
The overview structure of the proposed LLM4PR is illustrated in Figure \ref{fig:model_structure}. Specifically, $\text{QIA}_u$ takes query embedding $x_q$ and user feature embeddings $\mathcal{X}_u$ (\eg, id embedding, gender embedding, \etc.) as input, and produces a single embedding $e_u$ to represent the user. Similarly, $\text{QIA}_i$ takes query embedding $x_q$ and candidate/history item feature embeddings $\mathcal{X}_i$/$\mathcal{X}_h$ as input, and generates a hidden vector for each item in $\mathcal{I}$/$\mathcal{H}_u$. Note that, the query embedding and feature embeddings of the user/item fed into the QIA are fetched from previous stage (\ie, the ranking stage), instead of training from scratch. 
%Experiment results in Section \ref{sec:offline_exp} improve that the fetched embeddings, which are totally trained and contain meaningful information for ranking task, are beneficial to the re-ranking task. 
Subsequently, the user embedding $e_u$, candidate item embeddings $e_i$, behavior item embeddings $e_h$, and the raw query $q$ are combined using a post-ranking template $\mathcal{T}_{pr}^{main}$. The template provides a hybrid input instruction, containing both texts and embeddings, %for LLMs to re-rank the input candidates.
and the LLM directly predicts the final list in a generative manner according to the given  input instruction.
%For the given input instruction, the LLM directly predicts the ranking order in a generative manner . %In this paper, we employ capital letters  to represent the ranking order and decode the generated capital letter sequence to obtain the re-ranking result $\mathcal{I}^{\ast}$.
Next, we will introduce the designed QIA and training tasks for LLM4PR.

\subsection{Query-Instructed Adapter}\label{sec:qia}
% \begin{figure}[ht]
%     \centering
%     \includegraphics[width=0.90\columnwidth, right]{images/qia_target_att.drawio}
%     \vspace{-15pt}
%     \caption{Illustration of Query-Instructed Adapter (QIA). The QIA takes query embedding $x_q$ and user/item feature embeddings $\mathcal{X}$ as input and outputs a representation vector $e$ for the user/item.}
%     \label{fig:qia_struct}
%     \vspace{-10pt}
% \end{figure}
Traditional post-ranking methods utilize diverse features to represent a specific user or item. 
% For instance, user ID, gender, age, and statistical attributes are all effective features for users modeling. 
A straightforward way to utilize these features in LLMs is to combine all texts into one sentence as the input for LLM (serving as one baseline method in Section 5).  %which only takes texts as input, is converting them into nature languages and combined into one sentence as the input for LLM. 
However, considering current ranking methods include hundreds or even thousands of different features, 
% and the input for post-ranking model concurrently contains multiple items, 
the combined sentence may become excessively long, causing significant increase in inference time.
To this end, we propose the QIA component, which utilizes the attention mechanism to merge various features of the user/item and delivers a single embedding vector to represent the specific user/item.

% As shown in Figure \ref{fig:qia_struct}, 
Specifically, the QIA takes search query embedding $x_q\in\mathbb{R}^{1\times d_q}$ and multiple user/item features $\mathcal{X}=\{x_1,x_2,\cdots,x_k\}$ as input, where $x_j\in\mathbb{R}^{1\times d_j}$ is the $j$-th feature vector with dimension $d_j$.  We concatenate all the numerical features (\eg, category embeddings, predicted scores from the up-stream stage, \etc.) and treat it as an embedding vector. % which is also included in the $\mathcal{X}$. 
It is noted that the query embedding $x_q$ and feature vectors $\mathcal{X}$ are fetched from the ranking stage and fixed during the training phase. 
%For the query $q$, which is a nature language sentence, QIA will tokenize and embed it into token embeddings. The tokenizer and embedding layer used here are the same as the LLM. Then, an average pooling is applied on the query token embeddings, which obtains a hidden vector $x_q\in\mathbb{R}^{1\times d_{llm}}$ for the query, where $d_{llm}$ is the hidden size of the LLM. 
The search query $q$ reflects user demand and may align with specific characteristics of the items. For example, when a user searches for a \textit{romance film}, the model should focus on the genre of the movies.
%The search query $q$ carries user demand and may match certain characteristic of the items. For instance, when a user searches for \textit{romance film}, the model should focus more on the movie genre. 
Thus, we implement a feature field attention technique to automatically align various features based on the query.
Concretely, the projection heads are first applied on the query and feature embeddings to produce the Query $Q$, Key $K$ and Value $V$, which can be formulated as
\begin{gather}
    Q=x_qW_Q,\\
    K=\{x_1W_{K_1};x_2W_{K_2};\cdots;x_kW_{K_k}\}, \\
    V=\{x_1W_{V_1};x_2W_{V_2};\cdots;x_kW_{V_k}\},
\end{gather}
% \begin{equation}
%     \begin{aligned}
%          Q=x_qW_Q,\label{equ:atten_q}\\
%     K=\begin{pmatrix}
%         x_1W_{K_1}\\
%         x_2W_{K_2}\\
%         \vdots \\
%         x_kW_{K_k}
%     \end{pmatrix}, \\
%     V=\begin{pmatrix}
%         x_1W_{V_1}\\
%         x_2W_{V_2}\\
%         \vdots \\
%         x_kW_{V_k}
%     \end{pmatrix},
%     \end{aligned}
% \end{equation}
where $Q\in\mathbb{R}^{1\times d_h}$, $K,V\in\mathbb{R}^{k\times d_h}$, $W_Q\in\mathbb{R}^{d_{q}\times d_h}$, $W_{K_j},W_{V_j}\in\mathbb{R}^{d_j\times d_h}$, $d_h$ represents the attention hidden size, and $k$ is the number of feature fields. 
% For simplification, the key projection weight and value projection weights are denoted as $\mathcal{W}_K=\{W_{K_1},W_{K_2},\cdots,W_{K_k}\}$ and $\mathcal{W}_V=\{W_{V_1},W_{V_2},\cdots,W_{V_k}\}$ in Figure \ref{fig:qia_struct}. 
Then, a standard target attention is employed to derive the hidden vector 
\begin{math}
h_w=\text{Softmax}(\frac{QK^{\mathsf{T}}}{\sqrt{d_h}})V,
\end{math}
% \begin{equation}
%     \begin{aligned}
%         h_w=\text{Softmax}(\frac{QK^{\mathsf{T}}}{\sqrt{d_h}})V,
%     \end{aligned}
% \end{equation}
where $h_w\in\mathbb{R}^{1\times d_h}$.
% the attention score $s$ is obtained by
% \begin{equation}
%     \begin{aligned}
%         s=\text{Softmax}(\frac{QK^{\mathsf{T}}}{\sqrt{d_h}}),
%     \end{aligned}
% \end{equation}
% where $s\in\mathbb{R}^{1\times k}$. Next, the attention score $s$ is multiplied with $V$ to produce the hidden feature vector $h_{w}$:
% \begin{equation}
%     \begin{aligned}
%         h_{w}=\text{Flatten}(s^{\mathsf{T}}\times V),
%     \end{aligned}
% \end{equation}
% where $h_{w}\in\mathbb{R}^{1\times kd_h}$. 
% $\text{Flatten}(\cdot)$ means reshaping the input matrix into a one-dimensional vector. 
% In our scenario, it is equivalent to concatenate the $k$ row feature vectors.
% Note that, in contrast to the conventional dot-product manner which reduces multiple feature vectors into one hidden vector by weighted sum, QIA acquires the hidden vector by, informally named, weighted concatenation. 
% Since the feature vectors input to QIA come from different feature channels, we replace the sum operation by concatenation to explicitly maintain the individual information from these diverse channels. 
% \textcolor{red}{Note that,  which is coarse-grained.}
Note that, the QIA attention mechanism provides a fine-grained interaction approach at the feature level, which enriches the expression of the user/item embedding. 
% As contrast, the query and user/item embeddings serve as the input for the LLM, which interact with each other implicitly at token level.
Finally, an additional linear projection maps the size of the hidden feature vector to the LLM hidden size:
\begin{math}
e=\text{Linear}(h_w),
\end{math}
% \begin{equation}
%     \begin{aligned}
%         e=\text{Linear}(h_w),
%     \end{aligned}
% \end{equation}
where $e\in\mathbb{R}^{1\times d_{llm}}$ is the final embedding to represent the specific user/item. $d_{llm}$ is the hidden size for the LLM.

\subsection{Feature Adaptation Step}
\begin{figure}[ht]
    \includegraphics[width=0.95\columnwidth, right]{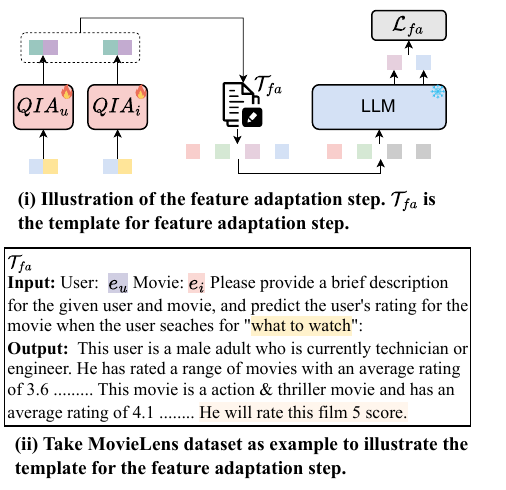}
    \vspace{-20pt}
    \caption{Illustration of the feature adaptation step. }
    \label{fig:1st_stage}
    % \vspace{-10pt}
\end{figure}
% (i) represents the training process for the feature adaptation stage which freezes the LLM and fine-tunes QIAs. Search query is omitted from the figure. (ii) demonstrates the template used in this stage.
LLMs are initially proposed to handle the NLP tasks, which take semantic texts as input. However, the user/item representations provided by the QIAs lack semantic information and cannot be understood by the LLM. Thus, we introduce a feature adaptation step to align the user/item representation embeddings with the LLM, serving as the preparatory phase for subsequent learning to post-rank. In this step, we freeze the LLM and fine-tune the QIAs to produce user/item representations that are most concordant with their semantic information.

% The QIAs are expected to produce the user/item representation that is most relevant to its semantic information, which can bridge the embedding distribution gap between the QIA and the LLM. 

%We named this Embedding Alignment. 
%Moreover, the feature embeddings (\ie, the input for QIAs) from the ranking stage contain abundant useful information for evaluating the user preferences for items, which serves as the prerequisite knowledge for reranking. Thus, we also expect to inject the useful information to the user/item representation (\ie, the output of QIAs and input for LLM) in this stage. We named this Knowledge Transfer.

% As shown in Figure \ref{fig:1st_stage}, in this stage, xxx generates descriptions for the given user and item, and predicts the user preferences for the item.

The process of feature adaptation step is illustrated in Figure \ref{fig:1st_stage}. The QIA takes query $q$ and user/item feature vectors ($\mathcal{X}_u$/$\mathcal{X}_i$) as input and produces the representation vector for the user/item ($e_u$/$e_i$). A template $\mathcal{T}_{fa}$ is introduced to integrate the query, user representation and item representation into a prompt sentence. $\mathcal{T}_{fa}$, demonstrated in Figure \ref{fig:1st_stage}(ii), is a hybrid template including both texts and numerical embeddings. There are two parts in the template, the input part is the input sentence for LLM and the output part is the target response that LLM is expected to produce. 

In template $\mathcal{T}_{fa}$, $e_u$ and $e_i$ are the user and item representations provided by $\text{QIA}_u$ and $\text{QIA}_i$, respectively. Besides the user and item information, $\mathcal{T}_{fa}$ also contains the instructions for LLM. Take MovieLens\cite{harper1025movielens} dataset as an example, $\mathcal{T}_{fa}$ requires the LLM to predict the user rating for the movie by providing descriptions for the input user and movie. 
% The responses for these two instructions are in the output part of $\mathcal{T}_{fa}$. 
The description generation aims to align the user/movie representation vector to its corresponding textual semantic. In this sense, the output of this instruction is constructed by organizing the various features of user/movie into a coherent sentence. Particularly, for categorical features such as gender, age group and occupation, we assemble the original meaning of the features as the description (\eg, this user is a \textit{male adult} who is currently \textit{technician or engineer}). 
For numerical features, which typically are the predicted values from the previous stage or statistics features like average rating scores, we directly articulate these features by combining their semantic interpretations and numerical values
%we describe these features with their feature meanings and numerical values straightforwardly 
(\eg, the user has rated a range of movies with an \textit{average rating of 3.6}). The user rating prediction instruction is designed to transfer the preference information from user/movie feature embeddings to its representation vector.
% which provides the priori knowledge for reranking.
%The user preference is measured by rating score in the MovieLens dataset, and in other scenarios, it could be other indicators like click.
% the indicator could be the video play duration or the click of the like button. 
%In the MovieLens dataset, the user preference is measured by rating score, and 
%In the output part of $\mathcal{T}_{fa}$, the response for the user rating prediction is the straightforward rating score, like \textit{he will rate this film 5 score}.
In the output part of $\mathcal{T}_{fa}$, the response for predicting user ratings is straightforwardly given as the rating score like 
\textit{he will rate this film 5 score}.

When constructing the input and output sentences for the LLM, we can train LLM4PR with such sentence pairs. Specifically, we freeze the LLM and train the parameters of the QIAs with the target of generating corresponding output sentence. We achieve this by maximizing the log likelihood for predicting target responses, and the loss function can be formulated as
\begin{equation}
    \begin{aligned}
        \mathcal{L}_{fa}=-\sum_{t=1}^{|y|}\log(P(y_t|q,\mathcal{X}_u,\mathcal{X}_i,y_{<t})),
    \end{aligned}\label{equ:loss_fa}
\end{equation}
where $\mathcal{X}_u$ and $\mathcal{X}_i$ represent the user and item feature embeddings, respectively. $y_{<t}$ indicates the tokens predicted in the previous $t-1$ steps.

\begin{figure}[ht]
    \includegraphics[width=0.95\columnwidth, right]{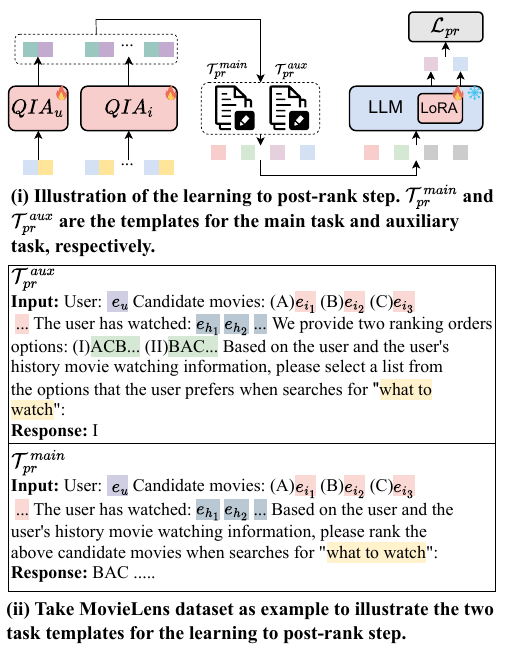}
    \vspace{-20pt}
    \caption{Illustration of the learning to post-rank step. }
    \label{fig:2nd_stage}
    % \vspace{-15pt}
\end{figure}
% (i) specifies the training process for the learning to re-rank stage which fine-tunes the QIAs and LLM with the two re-ranking tasks. Search query is omitted from the figure. (ii) demonstrates the auxiliary task and main task templates used in this step.

\subsection{Learning to Post-rank Step}
LLMs are originally designed for general purposes and necessitate proper modifications to effectively address the post-ranking task. Therefore, we introduce a learning to post-rank step that trains the generative LLM4PR framework. %The LLMs are designed for general purposes, requiring adjustments to motivate them to handle the re-ranking task. Thus, we introduce the learning to re-rank step, which trains LLM4PR to learn re-ranking in a generative manner.
We construct the input for the LLM and instruct it to generate a suitable order for candidate items, which serves as the main task. In addition, we further incorporate an auxiliary task that evaluates the qualities of two candidate lists via pairwise comparisons. The auxiliary task determines the superior list, and this auxiliary task serves as a simplified version for generating final post-ranking outcomes.

The procedure of learning to post-rank step is described in Figure \ref{fig:2nd_stage}, 
%$\text{QIA}_u$ takes query $q$ and user feature embeddings $\mathcal{X}_u$ as input and produces the user representation vector $e_u$. 
$\text{QIA}_u$ and $\text{QIA}_i$ produce the user representation $e_u$ and item representation vector $e_{i_j}$ for item $j$ in candidates $\mathcal{I}$, respectively. The representation vector for the $k$-th item in user's history behaviors is denoted as $e_{h_k}$.

%For each item in the candidates and history behavior list, $\text{QIA}_i$ takes query $q$ and item features as input and produces the corresponding item representation vector for each item. 
% The feature embeddings for the $j$-th item in candidates $\mathcal{I}$ and history behaviors $\mathcal{H}_u$ are represented as $\mathcal{X}_{i_j}=\{x^{i_j}_1,x^{i_j}_2,\cdots,x^{i_j}_{k_i}\}$ and $\mathcal{X}_{h_j}=\{x^{h_j}_1,x^{h_j}_2,\cdots,x^{h_j}_{k_i}\}$, respectively. 
% The corresponding representation vectors are indicated as $e_{i_j}$ and $e_{h_j}$, respectively.
%The representation vector for the $j$-th item in candidates $\mathcal{I}$ and history behaviors $\mathcal{H}_u$ are represented as $e_{i_j}$ and $e_{h_j}$, respectively.

Two hybrid templates $\mathcal{T}_{pr}^{main}$ and $\mathcal{T}_{pr}^{aux}$ are proposed to implement the learning to post-rank step. According to Figure \ref{fig:2nd_stage}(ii), the input parts of both templates are comprised of query $q$, user $e_u$, candidates $e_i$ and history behaviors $e_h$. 
In the auxiliary task template $\mathcal{T}_{pr}^{aux}$, two randomly generated candidate lists, denoted by the letters \textit{I} and \textit{II}, are provided for each training sample,  prompting the LLM to identify the superior option.
%it provides two ranking order options which are randomly generated for each training sample, and asks the LLM to select the superior one. 
%The two alternatives are denoted by the letters \textit{I} and \textit{II}. 
The quality of the generated ranking list is assessed by Normalized Discounted Cumulative Gain (NDCG). As a result, the target response indicates the list that achieves a higher NDCG value. 
Regarding the main task template $\mathcal{T}_{pr}^{main}$, it requires the LLM to post-rank all candidates. In this sense, the response to this instruction should be as plain as possible to simplify the generation process. Therefore, the output is formatted as a sequence of capital letters, with each letter symbolizing a specific item among the candidates. The final post-ranking list is determined by mapping letter \textit{A} to $i_1$ , letter \textit{B} to $i_2$ , and so forth.

%Thus, the output is designed as a sequence of capital letters, where each letter represents a corresponding item in the candidates. The final re-ranking list can be obtained by mapping letter \textit{A} to $i_1$, letter \textit{B} to $i_2$, and so on.

\iffalse
\textcolor{blue}{delete} The templates $\mathcal{T}_{pr}^{main}$ and $\mathcal{T}_{pr}^{aux}$ construct two training pairs for a single post-ranking training sample.
% These pairs are employed to train the LLM to acquire the re-ranking capabilities. 
During the training stage, the training pairs from these two tasks are combined to concurrently fine-tune the QIAs and LLM. 
\fi

When training the LLM4PR, the main task and auxiliary task are combined to concurrently fine-tune the QIAs and LLM. However, fine-tuning all the parameters of the LLM is time-consuming and requires massive computational resources. To this end, we utilize LoRA\cite{hu2022lora}, which brings a separate set of trainable parameters and keeps the pre-trained parameters of LLM frozen, to reduce the costs of fine-tuning.
In our implementation, LoRA parameters are exclusively integrated into the projection parts within the self-attention modules of the LLM.
We train the LLM4PR to predict the corresponding response using next-token prediction, which maximizes the log likelihood of the subsequent token. The loss function of this stage is represented as
\begin{equation}
    \begin{aligned}
        \mathcal{L}_{pr}=-\sum_{t=1}^{|y|}\log(P(y_t|q,\mathcal{X}_u,\mathcal{I},\mathcal{H}_u,y_{<t})),
    \end{aligned}\label{equ:loss_rerank}
\end{equation}
where $\mathcal{I}$ and $\mathcal{H}_u$ represent the set of candidate items and history behavior items, respectively. $y_{<t}$ indicates the tokens predicted in the previous $t-1$ steps.

\begin{algorithm}[t]
\caption{LLM4PR}\label{algo:algorithm}
\begin{flushleft}
\textbf{Input}: Parameters of $\text{QIA}_u$: $\theta_{QIA_u}$; Parameters of $\text{QIA}_i$: $\theta_{QIA_i}$; LoRA parameters: $\theta_{LoRA}$; Post-ranking dataset: $\mathcal{D}$; Maximum iteration step: $T_1$, $T_2$ \\
\textbf{Output}: Optimized learnable parameters
\end{flushleft}
\begin{algorithmic}[1]
\Statex // \texttt{feature adaptation step}
\For{$t=0$ to $T_1-1$}
    \For{sampled minibatch $\mathcal{D}_{batch}$ from $\mathcal{D}$}
        \State construct training data using $\mathcal{T}_{fa}$
        \State update $\theta_{QIA_u}$ and $\theta_{QIA_i}$ using Equation \ref{equ:loss_fa}
    \EndFor
\EndFor
\Statex // \texttt{learning to post-rank step}
\For{$t=0$ to $T_2-1$}
    \For{sampled minibatch $\mathcal{D}_{batch}$ from $\mathcal{D}$}
        \State construct training data using $\mathcal{T}_{pr}^{aux}$ and $\mathcal{T}_{pr}^{main}$
        \State update $\theta_{QIA_u}$, $\theta_{QIA_i}$, and $\theta_{LoRA}$ using Equation \ref{equ:loss_rerank}
    \EndFor
\EndFor
\State \Return  $\theta_{QIA_u}$, $\theta_{QIA_i}$, $\theta_{LoRA}$
\end{algorithmic}
\end{algorithm}
% \vspace{-10pt}

\subsection{Training and Inference}
In the training stage, the feature adaptation step and learning to post-rank step are performed sequentially, and the entire process for training LLM4PR is illustrated in Algo \ref{algo:algorithm}. In the inference stage, only $\mathcal{T}_{pr}^{main}$ is utilized to construct the inputs for the LLM, which instructs the LLM to generate the post-ranking list. The final post-ranking results are obtained by decoding the generated sequence produced by the LLM4PR. 

% \textcolor{red}{In industrial scenarios with strict time latency constraints,} 
% \textcolor{red}{we distill the LLM to smaller models, such as BART\cite{Lewis2020Bart}, to meet the requirements for online serving. Limited by the pages, this issue is not expanded upon in this paper.}

\section{Experiments}\label{sec:offline_exp}
% In this section, we firstly evaluate LLM4PR by comparing it with several state-of-the-art methods (SOTAs) on the re-ranking task. Then, to better understand the proposed LLM4PR, we conduct extensive ablation studies and analysis experiments.

\begin{table}[ht]
    \centering
    \caption{Statistics of the datasets.}
    \vspace{-10pt}
    \resizebox{0.95\columnwidth}{!}{
    \begin{tabular}{l|cccc}
        \toprule
        Dataset&\#Users&\#Items&\#Queries&\#Actions\\
        \midrule
        Covid&-&171,000&50&- \\
        NFCorpus&-&3,600&323&- \\
        Touche&-&382,000&49&- \\
        DBPedia&-&4,630,000&400&- \\
        SciFact&-&5,000&300&- \\
        Signal&-&2,860,000&97&- \\
        News&-&595,000&57&- \\
        Robust04&-&528,000&249&- \\
        MovieLens-1M&6,040&3,883&-&1,000,209 \\
        KuaiSAR(Search Part)&25,877&3,026,189&453,667&5,059,169\\
        \bottomrule
    \end{tabular}}
    \label{tab:dataset_statistic}
    \vspace{-10pt}
\end{table}

\subsection{Experiment Settings}
% For information retrieval dataset, the goal is to assess how well the LLM4PR can identify and post-rank relevant documents based on specific queries. For the search dataset, the focus is how the LLM4PR suggests items to users based on their interaction history and input queries.

\begin{table*}[ht]
    \centering
    \caption{Evaluation results (NDCG@10) on the  information retrieval task. \textbf{Bold} indicates the best results. The values in the table are presented as percentages (\%).}
    \vspace{-10pt}
    \resizebox{0.90\textwidth}{!}{
    \begin{tabular}{l|cccccccc|c}
        \toprule
        % Dataset$\rightarrow$&\multicolumn{4}{c|}{MovieLens-1M}&\multicolumn{6}{c}{KuaiSAR} \\
        % \cmidrule(lr){2-11}
        Method&Covid&NFCorpus&Touche&DBPedia&SciFact&Signal&News&Robust04&BEIR(Avg) \\
        \midrule
        BM25&59.47&30.75&\textbf{44.22}&31.80&67.89&33.05&39.52&40.70&43.42 \\
        monoBERT(340M)&70.01&36.88&31.75&41.87&71.36&31.44&44.62&49.35&47.16 \\
        monoT5(220M)&78.34&37.38&30.82&42.42&73.40&31.67&46.83&51.72&49.01 \\
        monoT5(3B)&80.71&\textbf{38.97}&32.41&44.45&\textbf{76.57}&32.55&48.49&56.71&51.36 \\

        Cohere Rerank-v2&81.81&36.36&32.51&42.51&74.44&29.60&47.59&50.78&49.45 \\
        LLM4PR(7B)&\textbf{83.75}&38.18&36.83&\textbf{46.35}&74.77&\textbf{33.86}&\textbf{49.85}&\textbf{57.28}&\textbf{52.61} \\
        \bottomrule
    \end{tabular}}
    \label{tab:main_results_ir}
    % \vspace{-10pt}
\end{table*}

\begin{table*}[ht]
    \centering
    \caption{Evaluation results on the search post-ranking task. \textbf{Bold} indicates the best results. The values in the table are presented as percentages (\%).}
    \vspace{-10pt}
    \resizebox{0.95\textwidth}{!}{
    \begin{tabular}{c|cccc|cccccc}
        \toprule
        Dataset$\rightarrow$&\multicolumn{4}{c|}{MovieLens-1M}&\multicolumn{6}{c}{KuaiSAR} \\
        \cmidrule(lr){2-11}
        % \midrule
        Model$\downarrow$&NDCG@3&NDCG@5&MRR@3&MRR@5&NDCG@3&NDCG@5&NDCG@10&MRR@3&MRR@5&MRR@10 \\
        \midrule
        LTR-DNN&63.30&77.59&42.71&50.14&63.52&68.80&83.06&43.33&47.32&50.05 \\
        LTR-DCN&70.21&80.91&50.92&56.18&64.33&69.37&83.84&45.35&49.37&51.75 \\
        DLCM&71.21&81.33&52.17&57.15&64.93&69.73&84.06&47.29&50.99&53.31\\
        Seq2Slate&73.47&82.18&55.15&59.14&65.17&70.19&84.22&47.40&51.02&53.46 \\
        SetRank&73.49&82.39&55.89&60.03&65.39&70.23&84.34&47.54&51.38&53.55 \\
        PRS&75.71&83.61&57.06&60.90&65.68&70.74&\textbf{84.96}&47.86&51.61&53.72 \\
        ChatGPT4RS&65.07&77.50&48.16&54.11&55.55&61.08&79.79&17.88&23.35&29.46 \\

        LLaRA&72.64&81.98&55.32&60.17&64.86&69.94&84.37&47.13&50.89&53.22 \\
        
        PTPR&71.63&81.43&57.60&61.63&65.27&70.05&84.58&47.35&50.57&53.08 \\
        LLM4PR(Ours)&\textbf{76.94}&\textbf{84.35}&\textbf{61.77}&\textbf{65.15}&\textbf{66.81}&\textbf{71.96}&84.91&\textbf{48.31}&\textbf{52.75}&\textbf{54.92} \\
        \bottomrule
    \end{tabular}}
    \label{tab:main_results}
    % \vspace{-10pt}
\end{table*}

\subsubsection{Datasets}
To validate the post-ranking performance of the proposed LLM4PR, we conduct experiments on two kinds of datasets: information retrieval dataset and search dataset.
% \wyh{claim KUSISAR is industrial dataset }
%To the best of our knowledge, there is no publicly available re-ranking dataset in search scenarios. 
For the information retrieval dataset, we select several sub-datasets from the BEIR dataset\cite{Thakur2021nips}, which contain only pure text feature. 
Moreover, we select MovieLens-1M\cite{harper1025movielens} and KuaiSAR\cite{sun2023kuaishousar} as the search datasets.
% For the search dataset, we conduct comprehensive experiments on two datasets: MovieLens-1M\cite{harper1025movielens} and KuaiSAR\cite{sun2023kuaishousar}.
Both datasets contain heterogeneous features and are modified to adapt the search post-ranking scenario. It is also noted that the KuaiSAR dataset is an industrial dataset collected from the Kuaishou App, the second largest short-video platform in China.
The statistics of the datasets are shown in Table \ref{tab:dataset_statistic}.

% We conduct comprehensive experiments on two datasets: MovieLens-1M\cite{harper1025movielens} and KuaiSAR\cite{sun2023kuaishousar}. Both datasets are modified to adapt the search re-ranking scenario, and the statistics of the datasets are shown in Table \ref{tab:dataset_statistic}.

% \noindent $\bullet$ \textbf{MovieLens-1M}. 
% The MovieLens-1M dataset, consisting of user ratings for movies, is a widely used dataset for evaluating ranking performance.
To adapt MovieLens-1M for the post-ranking scenario, following \cite{feng2021prs}, we sort movies in ascending order based on the timestamp for each user, and split the rating sequence into segments with a size of 5. 
% For segments with length smaller than 5, we pad the segment with random movies and set rating scores equal to 0.
%Then, we remove movies with duplicated rating scores from the segment and pad the segment with random movies to a fixed length of 5. The rating scores of the randomly padded movies are set to 0. 
% The movies in a segment serve as candidates. 
The target post-ranking list is the movies arranged in descending order based on their rating scores.
% The preceding movies for a certain segment are considered as the history behavior items.
Considering LLM4PR is designed the search scenario, we allocate a pseudo search query, such as \textit{what to watch} or \textit{movie recommendations}, for each segment.
For KuaiSAR dataset, we conduct experiments on the search part of KuaiSAR, treating the items in each search session as candidates for post-ranking. 
To leverage the textual information, which is encrypted in this dataset, we have contacted the authors and acquired the raw data.

% According to the timestamps, we divide the segments into an 8:1:1 ratio for training, validation, and testing, respectively.

% \noindent $\bullet$ \textbf{KuaiSAR}.
% KuaiSAR is a unified search and recommendation dataset containing genuine user behavior logs collected from the Kuaishou app. 
% We either truncate or pad each session to a fixed length of 10. The target order is ranking the clicked videos above the un-clicked ones. 

\begin{table*}[ht]
    \centering
    \caption{Ablation evaluation results on MovieLens-1M and KuaiSAR datasets. \textbf{Bold} indicates the best results. FA refers to the Feature Adaptation step and AT indicates the Auxiliary Task in the learning to post-rank step.}
    \vspace{-10pt}
    \resizebox{0.95\textwidth}{!}{
    \begin{tabular}{l|cccc|cccccc}
        \toprule
        Dataset$\rightarrow$&\multicolumn{4}{c|}{MovieLens-1M}&\multicolumn{6}{c}{KuaiSAR} \\
        \cmidrule(lr){2-11}
        % \midrule
        Model$\downarrow$&NDCG@3&NDCG@5&MRR@3&MRR@5&NDCG@3&NDCG@5&NDCG@10&MRR@3&MRR@5&MRR@10 \\
        \midrule
        LLM4PR&\textbf{76.94}&\textbf{84.35}&\textbf{61.77}&\textbf{65.15}&\textbf{66.81}&\textbf{71.96}&\textbf{84.91}&\textbf{48.31}&\textbf{52.75}&\textbf{54.92} \\
        \textit{w/o} QIA& 76.81&84.28&61.66&64.88&65.54&70.47&84.24&47.97&51.45&53.58\\
        \textit{w/o} FA&75.45&83.68&60.85&64.06&64.90&69.26&83.31&45.56&48.03&49.70 \\
        \textit{w/o} AT&75.98&84.07&61.31&64.28&65.05&69.67&83.78&47.20&50.48&52.84 \\
        \bottomrule
    \end{tabular}}
    \label{tab:ablation}
\end{table*}

\subsubsection{Baselines}
For the information retrieval dataset, we compare LLM4PR with several state-of-the-art (SOTA) passage ranking methods. The baselines include monoBERT\cite{DBLP:journals/corr/abs-1901-04085}, monoT5\cite{nogueira-etal-2020-document}, and Cohere Rerank\cite{Cohere-Rerank}. 
For the search dataset, we compare LLM4PR with LTR methods, conventional post-ranking methods, and LLM-based approaches. Specifically, we choose the LTR-DNN and LTR-DCN as the LTR baselines.
For conventional post-ranking methods, we select DLCM\cite{ai2018sigir}, Seq2Slate\cite{Bello2019seq}, SetRank\cite{Pang2020setrank}, and PRS\cite{feng2021prs} for comparisons.
For LLM-based benchmark, we select ChatGPT4RS\cite{dai2023llm4rs} and LLaRA\cite{LLaRA} as alternatives.
For ChatGPT4RS, we select the list-wise template to implement the post-ranking task. 
For LLaRA, we input all the candidate items into model, and the model chooses one item each time. Subsequently, the selected item is eliminated from candidates and the procedure is repeated for the remaining items until the ultimate list is acquired.
Furthermore, as discussed in Section \ref{sec:qia}, we combine all features into a single sentence as input for the LLM and train it using post-ranking targets. This alternative is referred to as `Pure Text Post-Ranking model' (PTPR).

\subsubsection{Evaluation Metrics}
For the information retrieval task, following \cite{sun2023listwisereranker}, we utilize NDCG@10 to evaluate the retrieval performance. For the search post-ranking task, we use NDCG@k and MRR@k as the evaluation metrics.
% The detailed descriptions of these two metrics are provided in Appendix \ref{sec:metric}.

\subsubsection{Implementation Details}
For the information retrieval dataset, following \cite{sun2023listwisereranker}, BM25 is used to retrieve the top 100 candidate passages. To simulate the post-ranking scenario, a pre-trained Sentence-BERT\cite{DBLP:conf/emnlp/ReimersG19} is used as the ranking model and obtain the passage embeddings. In the post-ranking stage, we utilize LLaMA2-7B\cite{touvron2023llama2} as the backbone for LLM4PR.
Following \cite{sun2023listwisereranker}, our LLM4PR model is trained on the MS MARCO dataset\cite{DBLP:conf/nips/NguyenRSGTMD16} and evaluated on the  sub-datasets of BEIR\cite{Thakur2021nips}.
% --
For the search dataset, the initial ranking lists and feature embeddings are produced by a DNN model\cite{paul2016dnn}, which serves as the ranking stage. The DNN leverages pre-trained BERT\cite{Devlin2019bert} to process the textual features.
We utilize LLaMA2-7B\cite{touvron2023llama2} as the backbone for LLM4PR on the MovieLens dataset and Baichuan2-7B\cite{yang2023baichuan} on the KuaiSAR dataset. 
For both tasks, The hidden size $d_h$ for QIA is set to 512.
We employ AdamW\cite{ilya2019adamw} with learning rate annealing from 1e-3 to 1e-6 to optimize the trainable parameters for both the feature adaptation step and the learning to post-rank step. 
Furthermore, we implement a linear warm-up strategy during the initial 10\% of all iterations. 
LoRA\cite{hu2022lora} is integrated into \texttt{[q\_proj,k\_proj]} and \texttt{[W\_pack]} modules for LLaMA2 and Baichuan2, respectively. The parameters for LoRA, including rank, alpha, and dropout, are configured as 32, 32, and 0.1, respectively. All experiments are conducted on machines equipped with 8$\times$ NVIDIA A100 GPUs, with a batch size set to 5 for each GPU. All experiments are repeated 10 times, and the average values are reported below.

\begin{table*}[ht]
    \centering
    \caption{Evaluation results of LLM4PR with feature embeddings of varying qualities. \textbf{Bold} indicates the best results. HQ, MQ, and PQ represent feature embeddings of High Quality, Medium Quality, and Poor Quality, respectively.}
    \vspace{-5pt}
    \resizebox{0.95\textwidth}{!}{
    \begin{tabular}{l|cccc|cccccc}
        \toprule
        Dataset$\rightarrow$&\multicolumn{4}{c|}{MovieLens-1M}&\multicolumn{6}{c}{KuaiSAR} \\
        \cmidrule(lr){2-11}
        % \midrule
        Model$\downarrow$&NDCG@3&NDCG@5&MRR@3&MRR@5&NDCG@3&NDCG@5&NDCG@10&MRR@3&MRR@5&MRR@10 \\
        \midrule
        LLM4PR-HQ&\textbf{76.94}&\textbf{84.35}&\textbf{61.77}&\textbf{65.15}&\textbf{66.81}&\textbf{71.96}&\textbf{84.91}&\textbf{48.31}&\textbf{52.75}&\textbf{54.92} \\
        LLM4PR-MQ&76.84&84.19&61.60&64.83&65.89&71.22&84.88&47.83&51.81&54.26 \\
        LLM4PR-PQ&76.43&83.96&61.23&64.53&65.27&70.05&84.58&47.35&50.57&53.08 \\
        \bottomrule
    \end{tabular}}
    \label{tab:emb_quality}
    % \vspace{-10pt}
\end{table*}

\subsection{Main Results}

For the information retrieval dataset, the evaluation results are reported in Table \ref{tab:main_results_ir}. As shown in the table, the proposed LLM4PR achieves the best average results (Best performances in 5 out of 8 sub-datasets, and ranking second in the remaining 3 sub-datasets), which fully demonstrate the superiority of the proposed method. Moreover, the model is trained with the MS MARCO dataset and evaluated on the BEIR dataset, verifying its remarkable transferability and generalization capability.

For the search dataset, the evaluation results are reported in Table \ref{tab:main_results}. We can see that, the proposed LLM4PR outperforms the baselines in most metrics in both datasets.
% demonstrating the superiority of our method. 
For the NDCG@10 metric in the KuaiSAR dataset, the performance of LLM4PR is also close to the best results of PRS.
% Specifically, for the MovieLens-1M dataset, LLM4PR achieves average improvements of 0.98\% and 4.48\% for the NDCG and MRR metrics, respectively, compared to the previous SOTA model. 
Moreover, the ChatGPT4RS merely feeds query and item texts into LLM without specific adaptation to the post-ranking task, leading to an inferior performance, indicating that generically pre-trained LLMs have limited capabilities for post-ranking. 
Additionally, the proposed LLM4PR achieves better results than other fine-tuned LLM-based methods, such as LLaRA and PTPR, demonstrating the superiority of our method.

% In contrast, our method realizes substantial enhancements over ChatGPT4RS, demonstrating the effectiveness of the proposed LLM4PR.
% \textcolor{red}{xxxxxxxxxxxxxxxxxxxxxxxxxxxxxxxxxxxx}

%Moreover, ChatGPT4RS merely feeds query and item texts into the LLM without specific adaptations for the re-ranking task, resulting in performance that falls short of traditional re-ranking methods. This suggests that generically pre-trained LLMs possess constrained abilities for re-ranking tasks. In contrast, our method realizes substantial enhancements over ChatGPT4RS, underscoring the efficacy of the proposed LLM4PR approach.

%pure LLMs, which are pre-trained for the general language tasks, have limited capabilities for re-ranking. Our method achieves significant improvements compared to the pure LLM, which demonstrates the effectiveness of the proposed LLM4PR.

%underperforms the traditional re-ranking methods, and even inferior to the LTR methods.

% For the KuaiSAR dataset, it even produces inferior performace compared to the LTR methods such as DNN.

%This indicates that pure LLMs, which are pre-trained for the general language tasks, have limited capabilities for re-ranking. Our method achieves significant improvements compared to the pure LLM, which demonstrates the effectiveness of the proposed LLM4PR.

\subsection{Ablation Studies}
In LLM4PR, we propose the QIA component and various tasks to address the post-ranking challenge. To explore the influence of these components in LLM4PR, we conduct ablation studies on the datasets and report the results in Table \ref{tab:ablation}. 
% Specifically, the FA refers to the Feature Adaptation step and AT indicates the Auxiliary Task in the learning to post-rank step.

%Given that the search queries in the MovieLens-1M dataset are simulated and convey limited information on user requirements, substituting the QIA has a negligible impact on overall performance.

\textit{QIA.} QIA utilizes the attention mechanism to combine the multiple feature embeddings. We substitute it with DNN and report the results as \textit{w/o} QIA in Table \ref{tab:ablation}. 
For the MovieLens-1M dataset, all metrics show a slight average decrease of 0.14\% in both NDCG and MRR. 
Since the search queries in MovieLens-1M dataset are synthetic and may convey limited information on user requirements, it is reasonable that replacing the QIA does not significantly affect the overall performance.
In contrast, the KuaiSAR dataset is a real-world dataset in the search scenario, encompassing authentic search queries for diverse requests. In this context, the decline extended to an average of 1.14\% for NDCG and 0.99\% for MRR, which verifies the importance of QIA and usefulness of integrating feature embeddings guided by the search query.

\textit{Feature Adaptation (FA).} The FA step is designed to align the user/item representation embeddings with the LLM. To evaluate the impact of this step, we exclude the FA step and train the LLM4PR directly on the post-ranking task. The results are presented as \textit{w/o} FA in Table \ref{tab:ablation}. As shown in the table, the performance of our model degrades significantly after removing the FA step. Specifically, in the MovieLens-1M dataset, the average NDCG decreases by 1.08\%, and the MRR decreases by 1.00\%. Similarly, the KuaiSAR dataset exhibits a notable average decrease of 2.07\% in NDCG and 4.23\% in MRR. %While in the KuaiSAR dataset, the NDCG and MRR are significantly decreased by an average of 1.17\% and 3.56\%, respectively. 
Therefore, removing the feature adaptation step notably affects the post-ranking performance, underscoring the indispensableness of FA step and confirming the significance of injecting semantic meanings into the representation embeddings through alignment with the LLM.

\textit{Auxiliary Task (AT).} The AT aims to equip the model with the ability to identify the superior list order. We assess its effectiveness by removing the auxiliary task from the training stage and report the results as \textit{w/o} AT in Table \ref{tab:ablation}. Particularly, the NDCG and MRR decrease for both datasets, which indicates that the auxiliary task positively contributes to enhancing the post-ranking performance.
% exhibit a decrease of 0.38\% and 0.36\%, respectively, for the MovieLens-1M dataset, and 0.69\% and 1.15\%, respectively, for the KuaiSAR dataset. 

\subsection{Further Analysis}
% High quality, medium quality, poor quality

\subsubsection{Feature Embeddings with Different Qualities} The LLM4PR utilizes the feature embeddings obtained from the previous stage (\ie, ranking stage). We explore how feature embeddings with different qualities influence the post-ranking performance. We extract embeddings from various stages of the training phase of the ranking model (\ie, the DNN model), denoted as Poor Quality (PQ, validation ranking AUC score of 0.5), Medium Quality (MQ, validation ranking AUC score of 0.7), and High Quality (HQ, validation ranking AUC score exceeding 0.85). We utilize these embeddings to train the LLM4PR and present the results in Table \ref{tab:emb_quality}. 
As shown in the table, the post-ranking performance consistently declines from high-quality to poor-quality embeddings, indicating that high-quality embeddings facilitate the learning to post-rank process for LLM4PR. We also find that even with poor-quality or medium-quality embeddings, %where the ranking AUC is 0.5 and the embeddings can be considered as random vectors,
LLM4PR is capable of delivering competitive results. In this context, the QIA component functions akin to an ``embedding layer''. The low-quality embeddings serve as the initialization, and during the training process, the QIA learns to map these low-quality vectors to meaningful embeddings for the LLM. Therefore, our model exhibits favorable compatibility for the input embeddings.

\begin{table}[ht]
    \centering
    \caption{Inference Time Cost. The Token column indicates the input token length and the Cost column represents the inference time.}
    \vspace{-10pt}
    \resizebox{0.80\columnwidth}{!}{
    \begin{tabular}{l|cc|cc}
        \toprule
        Dataset$\rightarrow$&\multicolumn{2}{c|}{MovieLens-1M}&\multicolumn{2}{c}{KuaiSAR} \\
        \cmidrule(lr){2-5}
        Model$\downarrow$&Tokens&Cost(ms)&Tokens&Cost(ms) \\
        % \toprule
        % &MovieLens-1M&KuaiSAR\\
        \midrule
        PTPR&772& 556&2135&1525 \\
        % LLaRA(n$\times$)&-& - \\
        LLM4PR&80&318&92&501 \\
        \bottomrule
    \end{tabular}}
    \label{tab:infer_time}
\end{table}

\subsubsection{Inference Time Cost}
% To evaluate the efficiency of the proposed LLM4PR, we present the inference time costs of various LLM-based methods in Table \ref{tab:infer_time}. 

% Since LLaRA outputs single item one time

As we mentioned in Section \ref{sec:qia}, directly concatenating all features into a single sentence as input for LLM  would significantly increase the inference time cost. 
To verify this statement, we compare the inference time costs of PTPR and LLM4PR, with the results presented in Table \ref{tab:infer_time}. 
As shown in the table, the input tokens for PTPR, which simply concatenates all features into a single sentence, are 9 and 23 times greater than those for LLM4PR in MovieLens and KuaiSAR datasets, respectively.
Consequently, the inference time cost for PTPR is 1.7 and 3.0 times higher than that of LLM4PR, respectively.
The increase in token length primarily comes from the numerical features. For instance, one movie item in the MovieLens dataset has a feature named `history average rating' with a value of 0.5. 
In LLM4PR, the rating value is converted into an embedding and serves as only one input token for the LLM. However, for PTPR, the same feature is input as plain text, `history average rating is 0.5', which will be tokenized into 8 tokens ('\texttt{\_history}', `\texttt{\_average}', `\texttt{\_rating}', `\texttt{\_is}', `\texttt{\_}', `\texttt{0}', `\texttt{.}', `\texttt{5}').
Thus, the proposed LLM4PR can efficiently processes diverse input features.

 \subsubsection{Backbone LLM with Different Sizes}
 \begin{figure}[t]
     \centering
     \includegraphics[width=\columnwidth]{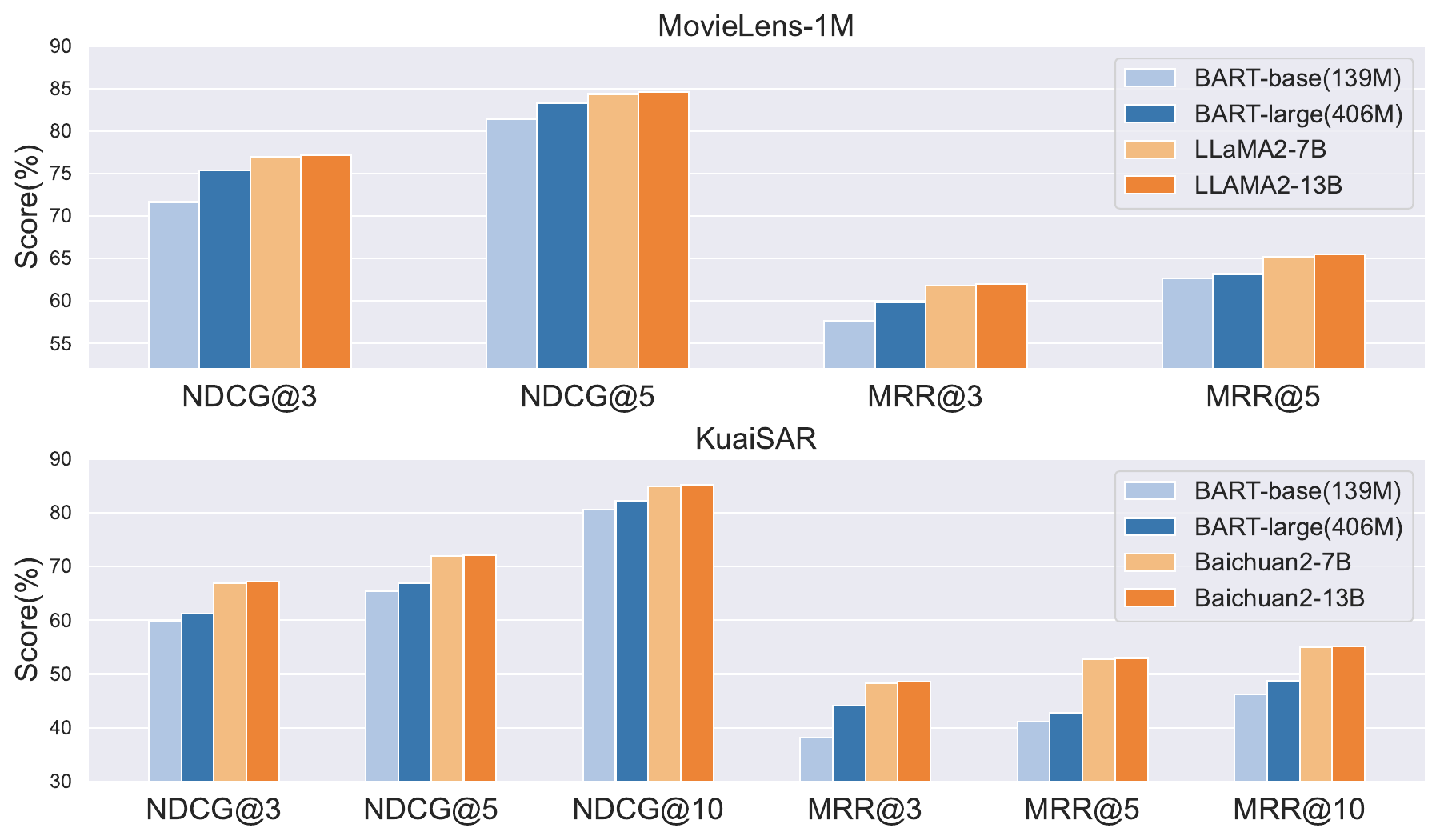}
     \vspace{-10pt}
     \caption{Post-ranking performance comparison \wrt \hspace{0.5mm} different sizes of backbone language models.}
     \label{fig:llm_size}
 \end{figure}

 To evaluate the impact of the backbone LLM size, we substitute the 7B LLM with different language models and provide the post-ranking performance in Figure \ref{fig:llm_size}. 
 As shown in this Figure, the post-ranking performances consistently improve with the increasing size of backbone models. This is consistent with the intuition that larger LLMs are inherently more powerful than smaller ones. However, when the model size reaches a sufficiently large scale, such as 7B, further enlarging the model size yields only marginal improvements. Therefore, to strike a balance between the training burden and overall performance, selecting a 7B version LLM as the backbone is a practical choice.

\subsubsection{Feature Adaptation Analysis}
\begin{figure}[ht]
    \centering
    \includegraphics[width=0.90\columnwidth]{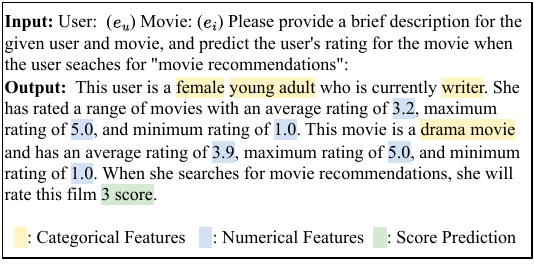}
    \vspace{-10pt}
    \caption{Example of the input and output from the feature adaptation step. Yellow, blue, and green indicate the predicted categorical features (\eg, gender, age, and occupation, \etc), numerical features (\eg, average rating scores, \etc), and rating score, respectively.}
    \label{fig:step1_example}
    % \vspace{-10pt}
\end{figure}

\begin{table}[ht]
    \centering
    \caption{Feature embedding alignment performance of the feature adaptation step for the MovieLens-1M dataset. Acc represents the accuracy.}
    \vspace{-10pt}
    \resizebox{0.70\columnwidth}{!}{
    \begin{tabular}{l|cc}
        \toprule
        Features&Acc (\%)&Auc (\%)\\
        \midrule
        Categorical Features&100& - \\
        Numerical Features&99.89&-\\
        Score Predictions &48.70&64.78\\
        \bottomrule
    \end{tabular}}
    \label{tab:step1_result}
    % \vspace{-10pt}
\end{table}

In this subsection, we validate the feature adaptation step and analyze its prediction results.
%to evaluate the \textcolor{red}{alignment performance}. 
%To validate the feature adaptation step, 
We utilize the template $\mathcal{T}_{fa}$ (shown in Figure \ref{fig:step1_example}) to generate the outputs, which include predicting categorical feature, numerical feature and rating scores. 
%to construct inputs and acquire the corresponding generation results. The generation example is presented in Figure \ref{fig:step1_example}, which includes predicting categorical feature, numerical feature and rating scores.
The generation results are evaluated accuracy and AUC values for the predicted features and rating scores, and the results are reported in Table \ref{tab:step1_result}. 
%To evaluate the generation results, we compute the accuracy and AUC values for the predicted features and rating scores, and the results are reported in Table \ref{tab:step1_result}. 
We observe that the LLM achieves exceptionally high accuracy scores for both categorical and numerical features, illustrating that the feature adaptation step successfully incorporates semantic meanings into the user/item representation vectors. Moreover, the LLM achieves 48.70\% accuracy score and 64.78\% AUC score for the rating score prediction, which is a 5-class classification task in MoiveLen-1M dataset. This demonstrates that the generated user/item representation vectors also encapsulate the user preference information, which is beneficial for learning to post-rank.
% \textcolor{blue}{ However, the score prediction results is not notable, demonstrating the score prediction in feature adaptation step is challenging for LLM.}

% We observe that, for both categorical and numerical features, the LLM achieves outstandingly high accuracy scores. This demonstrates that the feature adaptation step successfully injects the semantic meanings into the user/item representation vectors.
% Moreover, the LLM achieves 48.70\% accuracy score and 64.78\% AUC score for the rating score prediction-a 5-class classification task. This demonstrates that the generated user/item representation vectors also encapsulate the user preference information, which is beneficial for learning to re-rank.

% \subsubsection{LLM Frozen}
% 错误率，耗时，和原llama比较吧
% 不同size？!!!
% 不同base model 的影响

% few-shot的情况？也就是迁移能力，在一个数据集上训练，另一个预估！
% 两个step联合一起训练？

% \subsubsection{Infer Time Cost}

% \subsection{Online Results}
% \begin{table}[ht]
%     \centering
%     \caption{Online evaluation results. \textbf{Bold} highlights the most significant enhancement.}
%     \resizebox{0.95\columnwidth}{!}{
%     \begin{tabular}{l|cc}
%         \toprule
%         Models&Watch Time& Relevance Score \\
%         \midrule
%         Base Ranking Model& -& - \\
%         PRM \\
%         GE \\
%         LLM4PR-BART \\
%         \bottomrule
%     \end{tabular}}
%     \label{tab:online_result}
%     % \vspace{-10pt}
% \end{table}

\section{Conclusion}
In this paper, we introduce LLM4PR, the first LLM-based post-ranking framework designed for search engines. We introduce a QIA component to fuse heterogeneous features of users and items, and a feature adaptation step to align the user/item embeddings with the LLM. 
%Furthermore, we propose the feature adaptation step to align the user/item embeddings with the LLM. 
Furthermore, we leverage both the main and auxiliary tasks to fine-tune the model for learning post-ranking.
The extensive experiments on multiple datasets fully demonstrate the effectiveness and superiority of our proposed LLM4PR.

\bibliographystyle{ACM-Reference-Format}
\bibliography{llm4PR_wsdm}

% \appendix
% \section{Evaluation Metrics}\label{sec:metric}
% We use NDCG@k and MRR@k as the evaluation metrics. 
% Specifically, as shown in Equation \ref{equ:ndcg}, NDCG measures the quality of a ranked list by considering both the relevance and the position of each item. The score $s_i$ in Equation \ref{equ:ndcg} corresponds to the rating score for the MovieLens dataset. As for the KuaiSAR dataset, $s_i$ is 1 for clicked videos and 0 for un-clicked videos.
% MRR evaluates the effectiveness of a ranked list of items by focusing on the position of the first relevant item. The $rank_i$ in Equation \ref{equ:mrr} is the position of the first clicked item for KuaiSAR dataset. For MovieLens dataset, the $rank_i$ corresponds to the position of the item with the highest score within each segment. 
% Therefore, NDCG evaluates the overall ranking ability of the model, while MRR measures its capability to identify the best item. A higher value indicates superior performance for both metrics.

% \begin{equation}
%     \begin{aligned}
%         DCG@k=\sum_{i=1}^{k}\frac{2^{s_i}-1}{\log_2(i+1)} \\
%         NDCG@k=\frac{DCG@k}{IDCG@k}
%     \end{aligned}\label{equ:ndcg}
% \end{equation}
% \begin{equation}
%     \begin{aligned}
%         MRR@k=\frac{1}{N}\sum_{i=1}^N\frac{1}{rank_i}
%     \end{aligned}\label{equ:mrr}
% \end{equation}


\end{document}